\documentclass[useAMS,usenatbib]{mn2e}
\usepackage{amssymb}
\usepackage{journals}
\usepackage{graphicx}   % need for figures

\title[No Correlation Between Disc Scale-Height and Jet Power]{No Correlation Between Disc Scale-Height and Jet Power in GRMHD Simulations}
\author[P. C. Fragile, J. Wilson, \& M. Rodriguez]{P. Chris Fragile$^1$\thanks{KITP Visiting Scholar, Kavli Institute for Theoretical Physics, Santa Barbara, CA}, Julia Wilson$^1$ and Marco Rodriguez$^1$ \\
$^1$Department of Physics \& Astronomy, College of Charleston, Charleston, SC 29424, USA}
\begin{document}

\date{Released 2012 Xxxxx XX}

\pagerange{\pageref{firstpage}--\pageref{lastpage}} \pubyear{2012}

\maketitle

\label{firstpage}

\begin{abstract}
It is now well established that changes in the X-ray spectral state of black hole low-mass X-ray binaries are correlated with changes in the radio properties of those systems.  Assuming radio power is a proxy for jet power, we can say that the jet is continuously present in the hard state and undetectable (and therefore weaker) in the soft state.  Since the different accretion states are also generally assumed to be associated with different disc geometries -- the hard state with a hot, thick flow, and the soft state with a cold, thin disc -- we investigate the possibility that these two phenomena are linked; i.e., that the difference in disc geometry is the cause of the difference in observed jet power.  We do this by comparing various measures of jet power in numerical simulations of accretion discs of differing temperatures and thicknesses.  We perform these simulations using the general relativistic magnetohydrodynamic code {\sc Cosmos++} and a newly added cooling function, which allows us to regulate the disc scale height $H/r$ at different radii.  We find no apparent correlation between the disc scale height and jet power whenever we normalize the latter by the mass accretion history of each simulation.  We attribute this result to the role that the ``corona'' plays in confining and accelerating the jet (our corona may also be considered a failed MHD ``wind'').  The properties of the corona do not vary significantly from one simulation to another, even though the scale heights of the discs vary by up to a factor of four.  If this holds true in nature, then it suggests that the correlation between spectral state and jet power must be attributable to some other property, possibly the topology of the magnetic field.  Alternatively, it could be that the corona disappears altogether in the soft state, which would be consistent with observations, but has so far not been seen in simulations.  
\end{abstract}


\section{Introduction}
\label{sec:introduction}

Black hole low-mass X-ray binaries (LMXBs) exhibit radio emission (associated with jets) whose properties change with the observed X-ray spectral and, to a less well determined extent, temporal properties of the accretion disc \citep{fender04,fender09}.  Multiple systems have now been observed to follow similar evolutionary tracks in X-ray hardness-intensity diagrams, with their radio properties correlating with the position of the source in this diagram: e.g. GX 339-4 \citep{belloni05}, XTE J1650-500 \citep{corbel04}, and Cygnus X-1 \citep{wilms06}.  The salient details are as follows:  Whenever the system is observed to be ``hard,'' there is a steady, compact, optically thick radio jet.  As the X-ray luminosity increases during an outburst, the radio luminosity likewise increases.  At some point, the X-ray luminosity stops increasing and the spectrum begins to soften.
%; the system now approaches what is known as the ``jet line.''  Whenever it crosses this line, a powerful, optically thin radio outburst is observed.  
The spectrum continues to soften until the hard component effectively disappears and a pure ``soft'' state is reached (a weak power-law component may still be observed in the soft state).  Radio emissions consistent with a jet have so far not been observed in this state, although the data cannot rule out an intermittent, optically thin, jet \citep{fender09}.  In one of the most extreme examples, deep observations of the persistent soft-state source, 4U 1957+11, suggest the soft-state radio luminosity must be at least 300 times weaker than the expected hard-state luminosity for the same source \citep{russell11}.  After reaching the soft state, the system generally fades while remaining soft.  As it fades, the disc component scales as $L \propto T^4$, indicating a constant emitter size.  Eventually the system returns to the hard state at a significantly lower X-ray luminosity than the initial hard-to-soft transition.

In this paper we focus on the jet dichotomy between the hard and soft states.  In order to structure our argument, we take the view that almost any model of jets from black hole accretion discs can be boiled down to four main ``ingredients:'' 1) an energy reservoir that can be tapped, likely the rotational energy of the black hole (although possibly the disc); 2) a magnetic field configuration capable of tapping that energy, which almost certainly means the black hole or inner disc must be threaded by a net poloidal magnetic flux; 3) a confining medium to help accelerate the jet and convert the magnetic energy carried in the Poynting flux into kinetic energy of the plasma; and 4) some mechanism to introduce mass into the jet since a purely magnetic jet could not radiate.  Note that the first two ingredients are necessary and {\em sufficient} for a jet (or at least an outflow) to exist, while the latter two are important for it to be observable as a jet.  Having recapitulated jet models in this way, we ask which of the ingredients are likely to change dramatically when the disc switches from one spectral state to the other.

%Confinement is required for the acceleration of the jet \citep{komissarov07,tchekhovskoy09}, and may come from a hot, thick ADAF-like component \citep{mckinney06} or from a disc wind  \citep{tchekhovskoy09}.  Confinement can be very efficient at accelerating a jet, with a confined jet converting $>50$\% of its Poynting flux into kinetic energy flux, while the conversion efficiency for unconfined jets is only a few percent \citep{komissarov07,tchekhovskoy09}. In the case of mass loading, the amount of mass transferred from the disc to the jet depends linearly on the scale height, i.e. $\dot{M}_\mathrm{jet} \propto (H/r) \dot{M}_\mathrm{disc}$ \citep{tout96}.  Therefore, a geometrically thicker disc should both confine and mass-load a jet more effectively than a thin disc, making its jet more powerful and luminous.  Provided the scale height changes by at least a factor of $\approx 20$ when going from the hard state to the soft, then this geometrical interpretation would be consistent with the current observational constraints, if this picture is correct. 

Since the changes to jet power we are concerned with are found within individual sources, they can only be attributable to properties that can change on time-scales of a few days, since that is the observed time-scale of state transitions \citep[see, for example, the review by ][]{mcclintock06}.  This eliminates black hole rotational energy (ingredient \#1) from consideration, since the mass and angular momentum of the black hole are essentially constant over such time-scales.  The next three ingredients all remain plausible, however.  

Numerical work by \citet{beckwith08} [and more recently \citet{mckinney12}] shows quite nicely that a change in magnetic field topology (ingredient \#2) from a poloidal to a toroidal configuration leads to dramatically weaker or transient jets.  This is because, with a toroidal field, the black hole is not fed a consistent poloidal flux, which it needs to build and maintain a strong jet.  Therefore, a global field change could explain the observed jet dichotomy if the disc always maintains a net poloidal flux around the black hole in the hard state, but not in the soft state.  An argument along these lines was made by \citet{igumenshchev09}.  However, this does not account for why the disc itself undergoes a significant change during the state transition, as in most respects in numerical simulations, the properties of the disc are the same regardless of magnetic field topology (magnetically driven turbulence within discs is largely insensitive to the net magnetic flux, unlike jets) \citep{beckwith08}.  This implies that something else besides the field topology must also change during the state transition in order to explain the different appearance of the disc.  In our opinion, this need for two seemingly unrelated changes to happen nearly simultaneously every time a state transition occurs presents a significant hurdle for this picture.  Therefore, for now, we also eliminate ingredient \#2 from consideration.  

That leaves ingredients \#3 and \#4 as candidates.  First we point out that these two ingredients could be linked: the same medium that provides confinement for the jet might also be responsible for mass loading it.  This would be especially true if there were some geometrically thick, centrifugally-supported structure surrounding the jet, perhaps something like an advection-dominated accretion flow (ADAF) \citep{narayan94}.  The truncated disc model \citep{esin97} argues that just such a structure should be present in the hard state.  Furthermore, according to this model, this structure is supposed to disappear in the soft state, to be replaced by a geometrically thin Shakura-Sunyaev disc \citep{shakura73}, which extends all the way down to the innermost stable circular orbit (ISCO).  

In this paper, we explore whether this same geometrical argument can explain why jets are observed in the hard state but not the soft.  Our basic premise is that the jet (in terms of ingredients \#1 and \#2) is actually present in both accretion states, but so much less luminous in the soft state (because ingredients \#3 and \#4 are missing) that it has yet to be observed.  It may even be that the ``jet'' (fast outflow with a narrow opening angle) converts into a ``wind'' (slow outflow with a wide opening angle) when the system transitions from the hard to the soft state.  Such a picture would be consistent with the recent discovery in LMXBs of wide-angle winds that are observed in the soft state but not the hard \citep{ponti_12}.

The recent of incorporation of a new cooling function into the computational astrophysics code {\sc Cosmos++} has presented us with an opportunity to test our idea numerically.  We perform a series of general relativistic magnetohydrodynamic (GRMHD) simulations of accretion discs that differ only in their scale heights.  We then measure the output jet power and look for a correlation between this and the disc thickness that would be analogous to the observed correlation in nature.  Since we carefully track the mass accretion history of each simulation and treat the backgrounds identically, we can be confident that any differences we observe are due solely to the geometry of each disc (or equivalently the disc temperature, which controls the geometry).  Numerical limitations in treating the mass loading of the jet prevent us from being completely quantitative with our results; nevertheless, we can robustly test our basic idea.

Since this is the first GRMHD study of jets from relatively thin discs ($H/r \lesssim 0.1$), the results should also be of some general interest.  Along with testing our premise about the jet dichotomy in LMXBs, this work also extends our understanding of jets from black hole accretion simulations.  Previous studies of jets from such simulations \citep[e.g][]{devilliers05,hawley06,mckinney06}, did not include cooling functions.  As a result, the scale height was rather large ($H/r \gtrsim 0.1$) and could not be varied independently of other parameters.  The only other study we know of that tested jet properties specifically as a function of disc thickness is \citet{moscibrodzka09}, although their discs were embedded inside an in-falling, quasi-spherical envelope and they controlled the disc thickness by adjusting the adiabatic index $\Gamma$ of their gas.  Our work is distinct from theirs in both respects.

Our paper is organized as follows:  In \S \ref{sec:methods} we describe the numerical scheme used in this work, including how the simulations were initialized and the diagnostics used.  In \S \ref{sec:results} we present our main results, comparing jet power in simulations with different disc scale heights.  We find that, contrary to our arguments, there is very little correlation between the output jet power and the disc scale height when one normalizes for the mass accretion history.  In \S \ref{sec:corona} we speculate why our numerical results did not support our hypothesis.  Finally, in \S \ref{sec:conclusions} we present some concluding thoughts.  Throughout this paper we use units such that \(GM=c=1\); thus, units of length and time are \(r_G=GM/c^2\) and \(t_G=GM/c^3\), respectively.

\section{Numerical Method}
\label{sec:methods}

We solve the following set of conservative, general relativistic MHD equations, including radiative cooling,
\begin{equation}
\partial_t D + \partial_i (DV^i) = 0 ~,  \label{eqn:de} 
\end{equation}
\begin{equation}
 \partial_t {\cal E} + \partial_i (-\sqrt{-g}~T^i_t) =
      -\sqrt{-g}~T^\kappa_\lambda~\Gamma^\lambda_{0\kappa} + \sqrt{-g}~\Lambda u_t ~,
    \label{eqn:en}
\end{equation}
\begin{equation}
 \partial_t {\cal S}_j + \partial_i (\sqrt{-g}~T^i_j) =
      \sqrt{-g}~T^\kappa_\lambda~\Gamma^\lambda_{j\kappa} - \sqrt{-g}~\Lambda u_j ~,
    \label{eqn:mom}
\end{equation}
\begin{equation}
 \partial_t \mathcal{B}^j + \partial_i (\mathcal{B}^j V^i - \mathcal{B}^i V^j) =
     0 ~, 
      \label{eqn:ind} 
\end{equation}
where $D=W\rho$ is the generalized fluid density, $W=\sqrt{-g} u^t$ is the generalized boost factor, $V^i=u^i/u^t$ is the transport velocity, $u^\mu = g^{\mu \nu} u_\nu$ is the fluid 4-velocity, $g_{\mu\nu}$ is the 4-metric, $g$ is the 4-metric determinant, 
\begin{eqnarray}
{\cal E} & = & -\sqrt{-g} T^t_t \nonumber \\
 & = & -W u_t \left(\rho h + 2 P_B \right) - \sqrt{-g}\left(P + P_B\right) + \sqrt{-g} b^t b_t
\end{eqnarray}
is the total energy density, $h = 1 + \epsilon + P/\rho$ is the specific enthalpy, $\epsilon$ is the specific internal energy, $P$ is the fluid pressure, $P_B$ is the magnetic pressure, $T^\kappa_\lambda$ is the stress-energy tensor, $\Lambda$ is the cooling function (described in detail in \S \ref{sec:cooling}), and
\begin{equation}
{\cal S}_\mu = \sqrt{-g} T^t_j = W u_j \left(\rho h + 2 P_B \right) - \sqrt{-g} b^t b_j
\end{equation}
is the covariant momentum density. With indices, $\Gamma$ indicates the geometric connection coefficients of the metric; without indices, $\Gamma$ is the adiabatic index.  For this work we use an ideal gas equation of state (EOS),
\begin{equation}
P = (\Gamma-1)\rho\epsilon ~,
\end{equation}
with $\Gamma = 5/3$.

There are multiple representations of the magnetic field in our equations: $b^\mu$ is the magnetic field measured by an observer comoving with the fluid, which can be defined in terms of the dual of the Faraday tensor $b^\mu \equiv u_\nu {^*F^{\mu\nu}}$, and $\mathcal{B}^j = \sqrt{-g} B^j$ is the boosted magnetic field 3-vector. The un-boosted magnetic field 3-vector $B^i = {^*F^{\mu i}}$ is related to the comoving field by 
\begin{equation}
B^i = u^t b^i - u^i b^t ~.
\end{equation}
The magnetic pressure is $P_B = b^2/2 = b^\mu b_\mu/2$. 
%Note that, unlike in previous versions of \emph{Cosmos++}, we have absorbed the factor of $\sqrt{4\pi}$ into the definition of the magnetic fields, so-called Lorentz-Heaviside units.

Equations (\ref{eqn:de}) -- (\ref{eqn:ind}) are solved using the new High Resolution Shock Capturing (HRSC) scheme recently added to the {\sc Cosmos++} computational astrophysics code \citep{fragile12}.

\subsection{Initialization}

All of our simulations start with identical initial conditions, consisting of a gas torus orbiting within a low density background around a Kerr black hole.  The free parameters that describe the simulations are the black-hole spin ($a_*=a/M = c J/G M^2=0.93$), where $J$ is the angular momentum of the black-hole, the inner radius of the torus ($r_{\rm in}=15 r_G$), the radius of the pressure maximum ($r_{\rm centre}=45 r_G$), and the power-law exponent ($q=1.54$) used in defining the specific angular momentum distribution (and thus influencing the initial thickness),
\begin{equation}
\ell = -u_\phi/u_t \propto \left( -\frac{g_{t \phi}+\ell g_{tt}}{\ell g_{\phi
\phi} + \ell^2 g_{t \phi}}\right)^{q/2-1} ~.
\end{equation}
We follow the procedure in \citet{chakrabarti85} to solve for the initial, hydrodynamic-equilibrium state of the torus.  

To this torus we add weak poloidal magnetic field loops in order to seed the magneto-rotational instability (MRI) \citep{balbus91}.  The non-zero spatial components are $\mathcal{B}^r = -
\partial_\vartheta A_\varphi$ and $\mathcal{B}^\vartheta =
\partial_r A_\varphi$, where
\begin{equation}
A_\varphi = \left\{ \begin{array}{lll}
          C(\rho-\rho_{\rm cut})^2 \sin \left[ 2 N \log (r/S) \right] & \mathrm{for} & \rho\ge\rho_{\rm cut}~, \\
          0                  & \mathrm{for} & \rho<\rho_{\rm cut}~,
         \end{array} \right.
\label{eq:torusb}
\end{equation}
$N=1$ is the number of field loop centres, and $S = 1.1 r_{\rm in}$ is a scaling parameter.  The parameter $\rho_{\rm cut}=0.5 \rho_{\rm max,0}$, where $\rho_{\rm max,0}$ is the initial density maximum within the torus, is used to keep the field a suitable distance inside the surface of the torus initially.  Using the constant $C$, the field is normalized such that initially $\beta_{\rm mag} =P/P_B \ge \beta_{\rm mag,0}=10$ throughout the torus. This is to ensure that the initial magnetic field is weak.

In the background region not specified by the torus solution, we set up a low density non-magnetic gas, with profiles $\rho = 10^{-4} \rho_{\rm max,0} r^{-2.7}$, $e = \rho\epsilon =  10^{-6} \rho_{\rm max,0} r^{-2.7}$, and $V^r = (g^{tr}/g^{tt})[1-(r_G/r)^4]$.  Numerical floors of the form $\rho_{\rm floor} = 10^{-7} \rho_{\rm max,0} r^{-2.7}$ and $e_{\rm floor} = 10^{-9} \rho_{\rm max,0} r^{-2.7}$ are placed on the density and internal energy density.  This density floor is sufficiently low and falls off fast enough radially that it is very rarely applied in the simulations, even inside the relatively evacuated jets.  Instead, the most commonly applied floor, and the source of matter inside the jets, is the floor applied on the ratio $(\rho + \rho\epsilon)/P_B$.  Whenever this quantity drops below 0.01, both $\rho$ and $\rho\epsilon$ are rescaled by a factor appropriate to maintain this ratio.  This procedure is necessary as the code can not handle more extreme magnetization.

Our simulations are run for 7.5 orbital periods, as calculated at the pressure maximum, $r_{\rm centre}$.  This corresponds to \(\sim14200 t_G\), as the orbital period is \(\sim1900 t_G\).  We consider three different target thicknesses ($H_\ast/r = 0.04$, 0.08, and 0.16), plus two models where the disc has both a thick ($H_\ast/r \approx 0.16$) and thin ($H_\ast/r \approx 0.04$) component with a specified transition radius $r_\mathrm{tr}$.  The names and parameters of each of the simulations are presented in Table \ref{tab:params}.  Note that simulation h16rAll has a target scale height ($H_\ast/r = 0.16$) that is thicker than the disc can achieve over the course of the simulation (the starting thickness is $H/r \approx 0.08$).  This was intentional, as we wanted to have one ``no-cool'' reference simulation.  It is important, though, that the non-disc material is treated consistently in all our simulations.

\begin{table*}
\label{tab:params}
 \centering
 \begin{minipage}{140mm}
  \caption{Simulation parameters}
  \begin{tabular}{@{}lccccc@{}}
  \hline
   Simulation & $H_\ast/r$ & $r_\mathrm{tr}/r_G$ & $H/r$\footnote{Average of profiles in Fig. \ref{fig:scale} from $r_\mathrm{ISCO}=2.10 r_G$ to $10 r_G$, the \\approximate light-cylinder radius.  Errors represent one standard deviation \\($1\sigma$) about this mean.} & $H/r$\footnote{Average of profiles in Fig. \ref{fig:scale} from $r_\mathrm{ISCO}=2.10 r_G$ to $40 r_G$ or $r_\mathrm{ISCO}$ to $r_\mathrm{tr}$ \\and $r_\mathrm{tr}$ to $40 r_G$.  Errors represent $1\sigma$ about this mean.} & $\eta$\footnote{Averaged over the final 2.5 orbits of the simulations, $9500 \le t/t_G \le 14200$.  \\Errors represent $1\sigma$ about this mean.} \\   
 \hline
h04rAll & 0.04 &  & $0.09\pm 0.02$ & $0.07\pm 0.03$ & $0.39\pm 0.11$ \\
h04r15 & 0.16 & 15 & $0.10\pm 0.02$ & $0.09\pm 0.02$ & $0.47\pm 0.08$ \\
& 0.04 & & & $0.04\pm 0.01$ &  \\
h04r30 & 0.16 & 30 & $0.10\pm 0.02$ & $0.11\pm 0.03$ & $0.48\pm 0.08$ \\
& 0.04 & & & $0.04\pm 0.01$ &  \\
h08rAll & 0.08 & & $0.06\pm 0.01$ & $0.06\pm 0.01$ & $0.60\pm 0.12$ \\
h16rAll & 0.16 & & $0.11\pm 0.02$ & $0.12\pm 0.02$ & $0.52\pm 0.14$ \\
\hline
\end{tabular}
\end{minipage}
\end{table*}

\subsection{Grid}

Our simulations are carried out in the modified Kerr-Schild coordinates ($t$, $x_1$, $x_2$, $\phi$) in 2.5 spatial dimensions (all three spatial components of vector quantities are evolved, although symmetry is assumed in the azimuthal direction).  The grid is uniformly spaced in coordinates $x_1$ and $x_2$, with 1024 and 512 zones, respectively.  All curvature, both real and coordinate, is handled via the metric terms in {\sc Cosmos++}.  The key then is the transformations between the grid coordinates $x_1$ and $x_2$ and the true spatial coordinates, in this case the Kerr-Schild $r$ and $\theta$ coordinates.  These transformations are modeled after \citet{mckinney06}.  The radial coordinate is given by 
\begin{equation}
r = R_1 + R_0 e^{x_1^{n_r}} ,
\end{equation}
where \(R_1\) controls the radial concentration of zones towards \(r_{\rm BH}\) and \(n_r\) affects the radial distance covered by each zone.  We choose \(R_1 = -3\), \(R_0 = r_{\rm BH}\), and \(n_r = 10\), allowing our grid to cover a large enough radial extent to study the disc at small scales and jet at large distances with a reasonable number of zones.  Because we use horizon-penetrating Kerr-Schild coordinates, we are able to place the inner radial boundary of the grid inside the black hole horizon, $r_\mathrm{BH} = r_G (1+\sqrt{1-a^2_*}) = 1.37 r_G$, thus isolating it from the physical domain of interest.  The radial range extends from \(0.9~r_\mathrm{BH}\) to \(10^4~r_G\).  We use outflow boundary conditions at both the inner and outer radial boundaries.

The polar angle transformation is given by \citep{mckinney06}
\begin{equation}
\theta = x_2 + \frac{1}{2}[1-h(r)]\sin(2x_2) ,
\end{equation}
where \(h(r)\) alters the angular concentration of grid zones, with values between 0 and 1 concentrating zones near the equatorial plane and values between 1 and 2 concentrating zones near the poles.  We use the expression
\begin{equation}
h(r) = 2 - (2 - Q_j)\left(\frac{r}{r_{0j}}\right)^{-g_j} ,
\end{equation}
with \(Q_j = 0.7\) and \(r_{0j} = 3\).  This concentrates grid zones near the equator at small radii and near the poles at larger radii.  The angular index,
\begin{equation}
g_j = n_j\left[\frac{1}{2} + \frac{1}{\pi}\arctan\left(\frac{r - r_{2j}}{r_{1j}}\right)\right] , 
\end{equation}
controls the transition between the inner and outer radial regions.  We choose \(r_{1j} = 20\), \(r_{2j} = 120\), and \(n_j = 0.3\).  
%We also performed one simulation with $r_{1j} = 40$ and $r_{2j} = 240$, just to confirm that our choices for these parameters did not significantly affect our results.  
The polar angle covers the full range $0 \le \theta \le \pi$, and we use reflecting boundary conditions along the poles.

The advantage of this grid construction is that it allows us to resolve both the disc and the jet.  Even for our thinnest discs, our resolution is sufficient to ensure that the fastest growing poloidal field MRI modes are well resolved
[$\lambda_\mathrm{MRI} \equiv 2\pi v_\mathrm{A}/\Omega \gtrsim 8 \Delta z$, where $v_\mathrm{A}$ is the Alfv\'en speed].  We also confirm that $\alpha_{\rm mag} = -g_{\phi\phi}b^r b^\phi/P_B \approx 0.3-0.4$, as expected \citep{hawley11}.  At large radii, because the grid zoning becomes nearly cylindrical around the pole, we are able to maintain a nearly constant resolution across the jet core, which also becomes nearly cylindrical at large distances.  We are also able to maintain a reasonable number of zones across the interface between the jet and the surrounding material, which can be the corona, failed MHD wind, or ``background'' gas.

\subsection{Cooling Function}
\label{sec:cooling}

Since the goal of this work is to study how jet power depends on disc thickness in GRMHD simulations, we need a way to control this parameter.  We use a modified form of the cooling function from \citet{noble09}, which adds a radiative loss term to equations (\ref{eqn:en}) and (\ref{eqn:mom}).  The version we use has the form 
\begin{eqnarray}
\label{eqn:cooling}
\Lambda=\Omega e \left\{
\begin{array}{lr}
0 & Y < 1 \\
Y-1 & 1 \le Y \le 2 \\
1 & Y > 2 
\end{array} \right. ~,
\end{eqnarray}
where \(\Omega(r)=M^{1/2}/(r^{3/2}+aM^{1/2})\) is the relativistic orbital frequency and $Y=(\Gamma-1)\epsilon/T_\ast$ is the ratio of the actual to the target temperature.  The target temperature is estimated from the target scale height as
\begin{equation}
\label{eqn:tartemp}
T_\ast=\frac{\pi}{2}\left[\frac{H_\ast}{r}r\Omega(r)\right]^2 ~.
\end{equation}
The key difference between our cooling function and the one from \citet{noble09} is that, in our form, the cooling function has a maximum value of $\Lambda \le \Omega e$, whereas in \citet{noble09}, the cooling continues to be proportional to some power of $Y$, meaning the cooling rate can be quite high for gas that has a temperature far from the target value.  This is an important distinction; in the case where $\Lambda \propto Y^q$, the gas in the jets cools quite strongly, {\em and} that cooling would vary from one simulation to the next as we change the target scale height.  Since our goal is to compare jet power, these artificial variations are unacceptable.  In our version, the background gas and jets are largely unaffected by the cooling function, and only the disc is significantly altered.

As one final modification to the cooling function, we allow the target scale height to vary with radius.  In this work we consider either a single target scale height or two target scale heights, one that applies at small radii (and allows the disc to be thick) and one that applies at large radii (and forces the disc to be thin).  The purpose of having two target scale heights is to mimic the truncated disc model, where a thin accretion disc at large scales is truncated at some radius, inside of which the flow is significantly hotter and thicker.  Note that, in either case, the cooling of non-disc gas is the same, so our jets should not be affected directly by this difference.

Since it is conceivable that the cooling time of the gas, $t_\mathrm{cool} = e/\Lambda$, could sometimes be shorter than the MHD timestep, $\Delta t_\mathrm{MHD} \sim \Delta x/V$, where $\Delta x$ and $V$ are the characteristic zone length and velocity, respectively, we allow the cooling routine to operate on its own shorter timestep (subcycle), if necessary.  To prevent runaway loops inside the cooling function, we limit the cooling subcycle to 5 iterations, with a maximum change to the specific internal energy in each subcycle of $\Delta \epsilon = 0.1 \epsilon$.  

As a final note on the cooling function, we emphasize that it is not intended to capture any particular physical process.  It simply acts as a heat sink mechanism, allowing the disc to maintain a specified thickness.  Thus, it gives us very good control over the main variable we want to test.  Another nice feature is that, because there is no physical scale associated with this cooling function, our results can apply to black holes of any mass.

\subsection{Diagnostics}

To measure the scale heights of our discs, we use the following density-weighted expression 
\begin{equation}
\label{eqn:height}
\frac{H}{r} = \left(\frac{\int\sqrt{-g}\rho^2(\theta-\pi/2)^2\, d\theta d\phi}{\int\sqrt{-g}\rho^2\, d\theta d\phi\,}\right)^{1/2} ~.
\end{equation}
Using this diagnostic, we find that the discs in our simulations generally match their target scale heights quite well (as shown in Table \ref{tab:params} and discussed in \S \ref{sec:scale}).  
%If we instead weight our scale heights by $\rho^2$, as suggested by \citet{penna11}, then the measured scale heights are significantly ($\sim XX$\%) smaller than our target values.

%We define two different ``jets'' in our simulations.  For the first, which we refer to as the ``kinetic'' jet, we consider all material that is both unbound, \(-u_t>1\), and outbound, \(u^r>0\).  As Fig. \ref{fig:unbound} illustrates, this definition overlaps well with 
To define our ``jets'' we used the criterion $b^2/\rho > 1$, emphasizing that these are magnetically-driven jets \citep[see also][]{mckinney04,devilliers05}, powered by the Blandford-Znajek (BZ) mechanism \citep{blandford77}.  However, we wish to emphasize that, at this time, it is not clear exactly how our numerical jets might correspond to what observers are able to measure.  \citet{narayan11} argue that the ballistic jet observed in the steep power-law state is the one associated with the BZ mechanism \citep[but see][]{fender10}.  However, \citet{narayan11} fail to explain why the BZ jet would only be seen in one accretion state.  In our picture, the jet is expected to be present in all accretion states, since the fundamental ingredients (rotation plus open magnetic field lines) are argued to be present in all states.  The question is then shifted to whether or not the BZ jet is observable in all states.
%We argue then that the failure to observe the jet in the soft state is owing to its weakness, not its absence.

%Because of the large outer radius used in these simulations, the jet remains contained within the simulation domain for nearly the entire run.  Therefore, we are able to use global diagnostics, such as the total fluid and magnetic energies carried by the jet.  The kinetic and magnetic energy of the jet are defined as
%\begin{equation}
%\mathrm{KE}_{\rm j}(t) = \int \left( \sqrt{-g} T^{00}_{(FL)} - D - \Gamma E \right) dr d\theta d\phi
%\label{eqn:jetKE}
%\end{equation}
%and
%\begin{equation}
%\mathrm{ME}_{\rm j}(t) = \int \sqrt{-g} T^{00}_{(EM)} dr d\theta d\phi ~,
%\label{eqn:jetME}
%\end{equation}
We use a few different measures of jet power or efficiency in this work.  The first is the same measure of jet efficiency as \citet{tchekhovskoy11}, namely $\eta = (\dot{M} - \dot{E})/\langle\dot{M}\rangle$, where 
\begin{equation}
\dot{M} (t) = -\int \sqrt{-g} \rho u^r d\theta d\phi
\label{eqn:massFlux}
\end{equation}
is the mass accretion rate and
\begin{equation}
\dot{E} (t) = \int \sqrt{-g} T_t^r d\theta d\phi~
\label{eqn:energyFlux}
\end{equation}
is the energy flux, both taken at the black hole event horizon \(r_\mathrm{BH}\).  Angle brackets indicate a time-averaged quantity.  The negative sign in equation (\ref{eqn:massFlux}) indicates that the mass flux is positive when rest mass is transported into the black hole.  Similarly, $\dot{E}$ is constructed such that a positive value indicates a net flux of energy into the black hole.  

A more global measure of jet power can be made by summing the energy carried in the electromagnetic and fluid components of the jet ($b^2/\rho > 1$)
\begin{equation}
E (t) = \int \sqrt{-g} T_t^t dr d\theta d\phi~,
\label{eqn:jetEnergy}
\end{equation}
where $T^t_{t,{\rm FL}} = \rho h u^t u_t + P$ and $T^t_{t,{\rm EM}} = 2P_B u^t u_t + P_B - b^t b_t$.  We also track the energy flux $\dot{E}$ separately for the fluid and electromagnetic components along the length of the jet.  This is useful for tracking the conversion of magnetic energy into kinetic.  Finally we track the density-weighted Lorentz factor 
\begin{equation}
\label{eqn:lorentz}
\gamma = \frac{\int\sqrt{-g}\rho\sqrt{-g_{tt}}u^t\, d\theta' d\phi'}{\int\sqrt{-g}\rho\, d\theta' d\phi'\,} ~,
\end{equation}
where $d\theta'$ and $d\phi'$ are restricted to those parts of the radial shell that contain jet material ($b^2/\rho > 1$).

We also need to define a ``corona'' in our simulations.  Without doing more detailed radiative modeling, it is difficult to know for certain what component of our simulations might correspond to the corona observed in the hard-state of LMXBs.  For the purpose of this paper, we use the working definition that the corona is comprised of all of the bound ($-u_t < 1$) material surrounding the disc.  Most of this material has $\beta_\mathrm{mag} \le 1$.  It was originally part of the torus, but gets lifted out of the disc by magneto-centrifugal forces.  For these reasons it might be more appropriate to call our corona a failed MHD wind.

\section{Results}
\label{sec:results}

\subsection{Scale Height}
\label{sec:scale}

\begin{figure}
\includegraphics[scale=0.33]{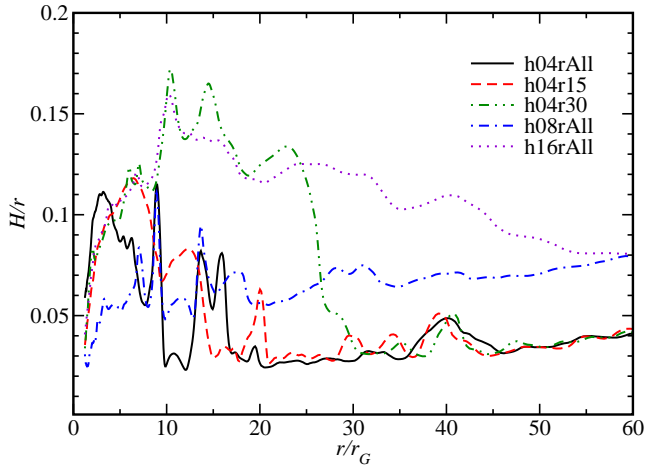}
\caption{Disc scale height \(H/r\) as a function of radius, averaged over the final 2.5 orbits, for each of our models.}
\label{fig:scale}
\end{figure}

Fig. \ref{fig:scale} shows the scale height $\langle H/r \rangle$ as a function of radius, averaged over the final 2.5 orbits of each simulation.  Even with time averaging, there is some variability in the scale height; this mostly represents regions where the cooling function is unable to keep up with local heating (near the event horizon) or where there are gaps in the disc such that the scale height is effectively measuring the corona and not the disc (around $15 r_G$ in a few simulations).  Particularly in simulation h04rAll, the disc breaks up into distinct clumps (rings) of matter, causing the scale height to appear large in the intervening gaps.  Nevertheless, the geometry in each simulation is largely as intended (see Table \ref{tab:params}), including the cases where the scale height changes from a thick to a thin component (simulations h04r15 and h04r30).  

\subsection{Magnetic Jet}

\begin{figure}
\includegraphics[scale=0.33]{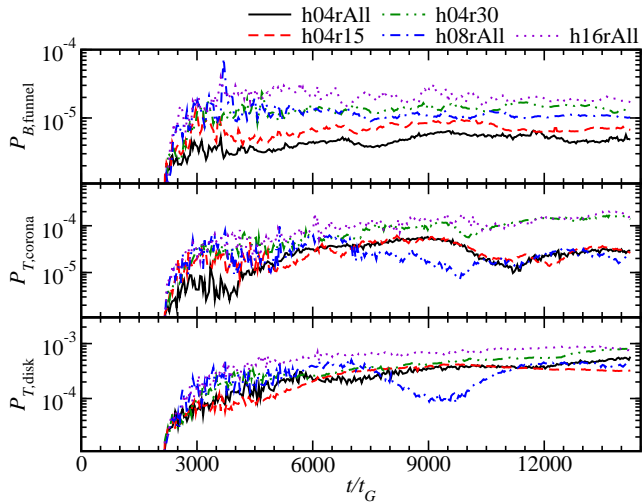}
\caption{Plot of the magnetic pressure in the funnel ($r=6 r_G$, $\theta \approx 0$), plus the total pressure in the corona ($r= 6 r_G$, $\theta = \pi/3$), and disk ($r=6r_G$, $\theta = \pi/2$), all in arbitrary units.}
\label{fig:pressure}
\end{figure}

\begin{figure}
\includegraphics[scale=0.33]{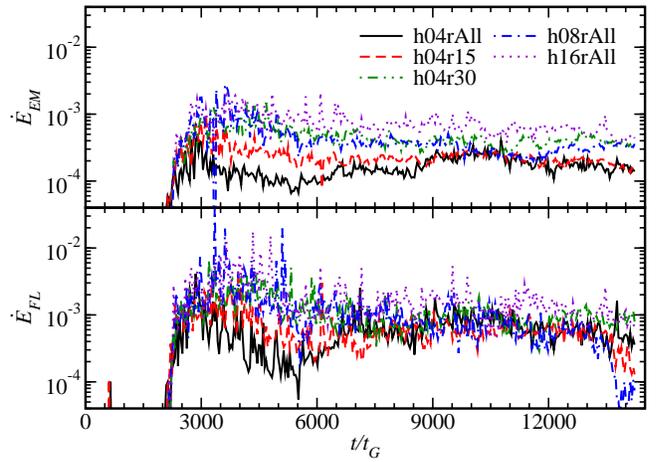}
\caption{Plot of the electromagnetic ($\dot{E}_{\rm EM}$) and fluid ($\dot{E}_{\rm FL}$) energy fluxes, both in arbitrary units, at the black hole event horizon.}
\label{fig:horizon_fluxes}
\end{figure}

In the Blandford-Znajek (BZ) model \citep{blandford77}, the power of the jet is predicted to depend on the angular velocity of the footpoint of the magnetic field and the strength of the poloidal magnetic field component, $P_\mathrm{BZ} \propto \Omega^2 b^2_p$.  The magnetic field is assumed to thread the event horizon, so the angular velocity is $\Omega_\mathrm{BH} = a/2r_\mathrm{BH}$.  Since $a$ (and therefore $\Omega_\mathrm{BH}$) is the same in all our simulations, we expect the Blandford-Znajek power of our jets should only vary if $b_p$ differs.  Fig. \ref{fig:pressure} shows that, indeed, the magnetic pressure inside the funnel, which close to the pole is dominated by the $b_p$ component, does vary proportionally with $H/r$.  This dependence is also reflected in the electromagnetic flux at the event horizon, $\dot{E}_{\rm EM} (r=r_\mathrm{BH})$, as shown in Fig. \ref{fig:horizon_fluxes}.   

\begin{figure}
\includegraphics[scale=0.33]{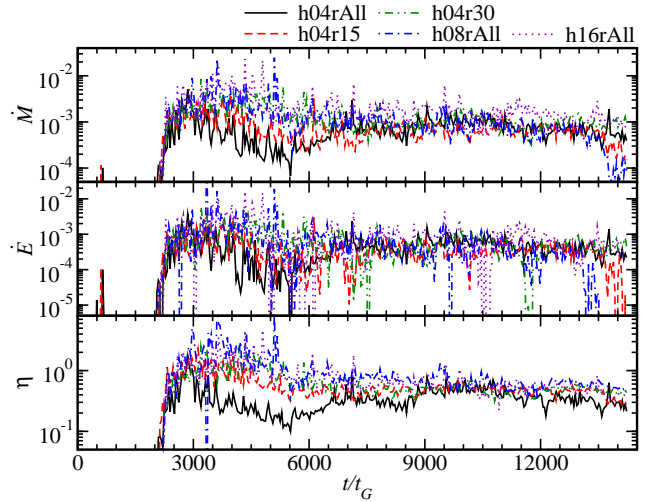}
\caption{Plot of the mass flux $\dot{M}$ and energy flux $\dot{E}$, both in arbitrary units, at the black hole event horizon and the jet efficiency $\eta = (\dot{M} - \dot{E})/\langle\dot{M}\rangle$.}
\label{fig:efficiency}
\end{figure}

\begin{figure}
\includegraphics[scale=0.33]{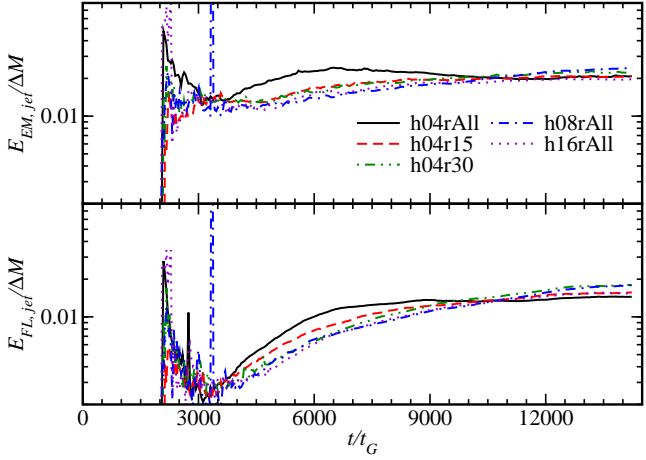}
\caption{Plot of the total energy carried in the electromagnetic and fluid components of the jet, both in arbitrary units, normalized by the amount of mass accreted onto the black hole.}
\label{fig:jet_energy}
\end{figure}

However, if we account for the fluid flux and normalize by $\dot{M}$, i.e. we plot a jet efficiency, then the dependence on $H/r$ disappears, certainly at late time, as shown in Fig. \ref{fig:efficiency} (see also Table \ref{tab:params}); all of our simulations settle to statistically equivalent jet efficiencies of $\eta \approx 0.5$.  The same independence is seen if we consider globally integrated measures of jet energy normalized by the total accreted mass.  For example, $E_{\rm EM}/\Delta M$ and $E_{\rm FL}/\Delta M$ show no discernible trend with $H/r$ in Fig. \ref{fig:jet_energy}.   

%Perhaps more importantly, the jet efficiencies in Fig. \ref{fig:horizon_fluxes} show no systematic variation with disc scale height.  This contradicts a claim in \citet{tchekhovskoy12} that thicker discs yield higher efficiencies.  However, in that work, the discs are all in the ``magnetically-arrested'' state, in which accretion onto the black hole is restricted by an accumulation of very large magnetic fluxes in the vicinity of the hole.  When the disc is in such a state, the amount of magnetic flux that threads the black hole depends not just the mass accretion rate, but also on the scale height of the disc \citep{mckinney12}.  This is a very different state than what we are exploring.  We also point out that \citet{tchekhovskoy12} consider much thicker discs ($H/r > 0.3$) than we do, which may also contribute to the differing conclusions.  

\subsection{Energy Conversion}

\begin{figure}
\includegraphics[scale=0.4]{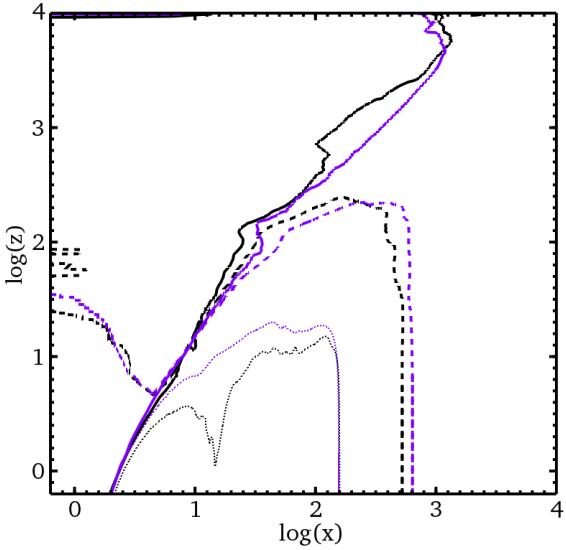}
\caption{Contours of the magnetically-dominated jet boundary, $b^2/\rho = 1$ ({\em solid lines}), the corona, $-u_t < 1$ ({\em dashed lines}), and the approximate disc surface, $\rho = 0.01\rho_{\rm max,0}$ ({\em thin, dotted lines}), for simulations h04rAll ({\em black}) and h16rAll ({\em purple}).  All quantities are time-averaged over the final 2 orbits.
\label{fig:components}}
\end{figure}

\begin{figure}
\includegraphics[scale=0.33]{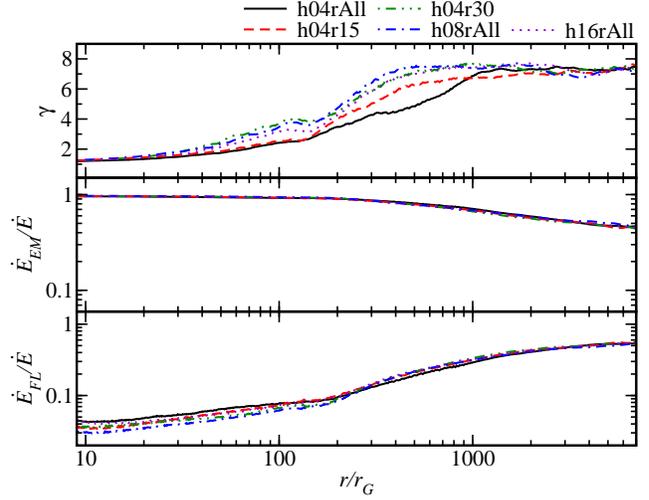}
\caption{Plot of the density-weighted shell average of the jet Lorentz factor $\gamma$ ({\em top}), the fraction of jet flux carried in the electromagnetic component $\dot{E}_{\rm EM}/(\dot{E}_{\rm EM} + \dot{E}_{\rm FL})$ ({\em middle}), and the fraction of jet flux carried in the fluid component $\dot{E}_{\rm FL}/(\dot{E}_{\rm EM} + \dot{E}_{\rm FL})$ ({\em bottom}).}
\label{fig:jet_fluxes}
\end{figure}

Having demonstrated that all of our jets contain the same power when normalized by the mass accretion history, we consider the question: How efficiently is this magnetic energy flux converted into kinetic energy flux within the jet?  Here we expect confinement to play a role, and it does, although not exactly in the way we anticipated.  Fig. \ref{fig:components} shows that the jet is confined out to a radius $r \approx 150 r_G$, not by the disc, but by what we label the corona.  Fig. \ref{fig:components} also illustrates that the size, or extent, of the corona is not strongly dependent on the scale height of the disc, since it appears nearly identical for our two most extreme simulations (h04rAll and h16rAll).  This is consistent with our intention in setting up the cooling function (\S \ref{sec:cooling}); only the disc, not the corona, should be affected by changes to the target scale height.  Fig. \ref{fig:jet_fluxes} then shows that the Lorentz factor indeed grows over the region of confinement, but the growth is modest, with $\gamma$ only reaching values of 3--4 by $r = 150 r_G$.

The rest of the acceleration actually occurs after the confinement from the corona cuts off ($r > 200 r_G$ in Fig. \ref{fig:jet_fluxes}).  The additional acceleration comes from the conversion of magnetic energy flux to kinetic, as can be seen in the bottom two panels of Fig. \ref{fig:jet_fluxes}.  This happens most efficiently once the matter passes through the outer fast magnetosonic surface \citep{eichler93,begelman94,tomimatsu94,mckinney06}.  Nevertheless, the modest acceleration provided by confinement is crucial, since it is only with this acceleration that the jet is able to achieve sufficient velocity to pass through the final critical surface and efficiently convert its energy to bulk kinetic motion.

Ultimately, in the outermost parts of our simulation domain ($r \gtrsim 5000 r_G$), we end up with a near equipartition between the magnetic and kinetic energy fluxes and a Lorentz factor of $\approx 7$, consistent with previous numerical studies \citep{devilliers05,hawley06,mckinney06}.  Most importantly, we again see there is very little difference in the behavior of the jets; certainly the variation is much smaller than the factor of $\approx 4$ difference in scale height.

%\begin{figure}
%\includegraphics[scale=0.4]{h04rAll_lorentz.ps}
%\includegraphics[scale=0.4]{h16rAll_lorentz.ps}
%\caption{Plot of the generalized Lorentz factor, $u^t = W/\sqrt{-g}$, IF POSSIBLE USE $\gamma$ for simulations h04rAll ({\em left}) and h16rAll ({\em right}) averaged over the final 2 orbits of the simulation. MAKE THIS SAME PLOT WITH $\vert V \vert$ TO CONFIRM IT BASICALLY APPEARS THE SAME.}
%\label{fig:lorentz}
%\end{figure}

%Fig. \ref{fig:jetke} shows the volume integrated kinetic and magnetic energies over the last \(\sim\)3.5 orbits, normalized by the mass accreted by the black hole.  SUMMARIZE RESULTS

%\begin{figure}
%\includegraphics[scale=0.33]{jetEnergy}
%\caption{Plot of total (volume-integrated) jet kinetic and magnetic energies, normalized by the total accreted mass.}
%\label{fig:jetke}
%\end{figure}

\section{Cold Corona}
\label{sec:corona}

Our two main arguments for why thicker discs might lead to more powerful observed jets were that thicker discs would presumably lead to more mass loading and would also better confine their jets.  However, the results in the previous section demonstrate two critical flaws in this argument: 1) confinement appears to be provided by the corona, not the disc; and 2) the bulk of the acceleration happens after the confinement ends.  Nevertheless, this apparent contradiction of our original hypothesis may point us in a more promising direction: if it is the corona that provides the confinement, then perhaps the critical aspect of the hard-to-soft transition that could explain the fading of the jet is the disappearance of the corona.  X-ray spectral observations already strongly suggest that while the hard state emission is dominated by a corona, the soft state spectrum shows little or no evidence for this component \citep{mcclintock06}.  If the corona indeed disappears in the soft state, then the only question that remains is if this also explains the fading of the jet.

Interestingly, the corona has never been observed to disappear in any GRMHD simulations that we are aware of, although there is some evidence for it being weaker in simulations that have no net poloidal flux \citep[see Fig. 6 of][]{beckwith08}.  In nature its disappearance may be related to some sort of thermal instability or change in ionization level, neither of which can be captured using the rudimentary treatments of gas thermodynamics currently implemented in global GRMHD disc simulations.  Nevertheless, we tried to create a smaller corona by conducting a few simulations that used the original \citet{noble09} cooling function, such that the non-disc gas cooled to a much greater extent than in the simulations presented here.  Even so, we still found no significant differences between the coronae of these alternative simulations and the ones shown in Fig. \ref{fig:components}.  This, perhaps, should not be surprising since these coronae are generated by magneto-centrifugal, not thermal, forces.  If the coronae in nature are similarly driven, then it may yet be that a change in magnetic field topology is required to explain both the disappearance of the jet {\em and} the corona.

\section{Conclusions}
\label{sec:conclusions}

The fundamental conclusion of this work is that there is no correlation between disc scale height and jet power in GRMHD simulations.  Varying the disc scale height by almost a factor of four ($0.04 \lesssim H/r \lesssim 0.16$) produces no meaningful differences in the measured jet efficiencies nor bulk Lorentz factors.  

Our main reason for suspecting that the disc scale height might affect jet power was that we presumed that the disc played an important role in providing confinement.  Based on our simulations, it appears this is not the case; rather, it is the corona that provides the confinement.  This finding contradicts the conjecture put forth in \citet{beckwith09} that it is the disk pressure that sets the level of (magnetic) pressure in the funnel.  Interestingly, we also find that the corona is not sensitive to the level of cooling applied in our simulations, and so, its properties do not vary appreciably from one simulation to the next.

In the Introduction, we suggested that jets ought to be present in all accretion states, since the necessary ingredients (rotation plus open magnetic field lines) seem likely to be present in all states.  We would then attribute the differing jet observations in each state to the degree of confinement provided.  In light of our results, we must modify this picture to either suggest that the different jet behavior between the hard and soft states must be due to a collapse of the corona or a change of field topology (scenarios that could not be explored using the methods of this paper).

As always, there is concern as to whether or not our simulations have run long enough to reach meaningful equilibrium states.  This is especially pertinent for our 2D simulations, as these can never truly reach a steady state.  The reason is that, after a period of initial vigorous growth, the MRI in 2D simulations steadily decays because the dynamo action that normally sustains it requires access to non-axisymmetric modes that are obviously inaccessible.  There are other pathologies related to the forced axisymmetry of 2D simulations, which may explain, for instance, the gaps that are observed in some of our disks in Figure \ref{fig:scale}.  

Another issue is that our simulations, like many previous ones \citep[e.g.][]{devilliers05,hawley06,mckinney09}, only supply a limited amount of magnetic flux to the black hole.  This restricts the range of jet efficiencies that we can hope to observe.  If, on the other hand, a very large amount of magnetic flux is available to the black hole, the black hole flux can saturate, with magnetic pressure regulating the accretion process, as demonstrated recently by \citet{igumenshchev03,igumenshchev08,tchekhovskoy11,mckinney12}.  In this ``magnetically choked'' state, the jet power does depend on disc scale height, as the thickness of the disc plays a role in regulating how much flux reaches the black hole \citep{tchekhovskoy12}.  Since our simulations only supply a limited amount of flux, we are not able to observe this effect.

Disc scale height may also play a role in regulating jet power by setting how much of the black hole horizon participates in the BZ mechanism; only that portion where $b^2/\rho > 1$ can power the jet.  In this case, it would only be the scale height very close to the event horizon that would matter.  Figure \ref{fig:components} indicates that this event horizon ``covering factor'' is nearly identical for our two most extreme simulations.  Thus, we do not expect to observe this effect, either.

Clearly these concerns leave room for further exploration of the topic of how disc scale height might effect jet power.  Our results also suggest that further study of the coronae in GRMHD simulations is warranted.

\section*{Acknowledgments}
We thank Kris Beckwith, Rob Fender, Sera Markoff, Jonathan McKinney, Alexander Tchekhovskoy, and the anonymous referee for useful comments on this manuscript.  This work was supported in part by the National Science Foundation under Grant No. NSF PHY11-25915 and by NSF Cooperative Agreement Number EPS-0919440 that included computing time on the Clemson University Palmetto Cluster.  MR gratefully acknowledges the support of Summer Undergraduate Research with Faculty (SURF) grant from the College of Charleston.

%\bibliographystyle{mn2e}
%\bibliography{refs}

\begin{thebibliography}{}

\bibitem[\protect\citeauthoryear{{Balbus} \& {Hawley}}{{Balbus} \&
  {Hawley}}{1991}]{balbus91}
{Balbus} S.~A.,  {Hawley} J.~F.,  1991, \apj, 376, 214

\bibitem[\protect\citeauthoryear{{Beckwith}, {Hawley} \& {Krolik}}{{Beckwith}
  et~al.}{2008}]{beckwith08}
{Beckwith} K.,  {Hawley} J.~F.,    {Krolik} J.~H.,  2008, \apj, 678, 1180

\bibitem[\protect\citeauthoryear{{Beckwith}, {Hawley} \& {Krolik}}{{Beckwith}
  et~al.}{2009}]{beckwith09}
{Beckwith} K.,  {Hawley} J.~F.,    {Krolik} J.~H.,  2009, \apj, 707, 428

\bibitem[\protect\citeauthoryear{{Begelman} \& {Li}}{{Begelman} \&
  {Li}}{1994}]{begelman94}
{Begelman} M.~C.,  {Li} Z.-Y.,  1994, \apj, 426, 269

\bibitem[\protect\citeauthoryear{{Belloni}, {Homan}, {Casella}, {van der Klis},
  {Nespoli}, {Lewin}, {Miller} \& {M{\'e}ndez}}{{Belloni}
  et~al.}{2005}]{belloni05}
{Belloni} T.,  {Homan} J.,  {Casella} P.,  {van der Klis} M.,  {Nespoli} E.,
  {Lewin} W.~H.~G.,  {Miller} J.~M.,    {M{\'e}ndez} M.,  2005, \aap, 440, 207

\bibitem[\protect\citeauthoryear{{Blandford} \& {Znajek}}{{Blandford} \&
  {Znajek}}{1977}]{blandford77}
{Blandford} R.~D.,  {Znajek} R.~L.,  1977, \mnras, 179, 433

\bibitem[\protect\citeauthoryear{{Chakrabarti}}{{Chakrabarti}}{1985}]{chakrabarti85}
{Chakrabarti} S.~K.,  1985, \apj, 288, 1

\bibitem[\protect\citeauthoryear{{Corbel}, {Fender}, {Tomsick}, {Tzioumis} \&
  {Tingay}}{{Corbel} et~al.}{2004}]{corbel04}
{Corbel} S.,  {Fender} R.~P.,  {Tomsick} J.~A.,  {Tzioumis} A.~K.,    {Tingay}
  S.,  2004, \apj, 617, 1272

\bibitem[\protect\citeauthoryear{{De Villiers}, {Hawley}, {Krolik} \&
  {Hirose}}{{De Villiers} et~al.}{2005}]{devilliers05}
{De Villiers} J.-P.,  {Hawley} J.~F.,  {Krolik} J.~H.,    {Hirose} S.,  2005,
  \apj, 620, 878

\bibitem[\protect\citeauthoryear{{Eichler}}{{Eichler}}{1993}]{eichler93}
{Eichler} D.,  1993, \apj, 419, 111

\bibitem[\protect\citeauthoryear{{Esin}, {McClintock} \& {Narayan}}{{Esin}
  et~al.}{1997}]{esin97}
{Esin} A.~A.,  {McClintock} J.~E.,    {Narayan} R.,  1997, \apj, 489, 865

\bibitem[\protect\citeauthoryear{{Fender}, {Belloni} \& {Gallo}}{{Fender}
  et~al.}{2004}]{fender04}
{Fender} R.~P.,  {Belloni} T.~M.,    {Gallo} E.,  2004, \mnras, 355, 1105

\bibitem[\protect\citeauthoryear{{Fender}, {Gallo} \& {Russell}}{{Fender}
  et~al.}{2010}]{fender10}
{Fender} R.~P.,  {Gallo} E.,    {Russell} D.,  2010, \mnras, 406, 1425

\bibitem[\protect\citeauthoryear{{Fender}, {Homan} \& {Belloni}}{{Fender}
  et~al.}{2009}]{fender09}
{Fender} R.~P.,  {Homan} J.,    {Belloni} T.~M.,  2009, \mnras, 396, 1370

\bibitem[\protect\citeauthoryear{Fragile, Gillespie, Monahan, Rodriguez \&
  Anninos}{Fragile et~al.}{2012}]{fragile12}
Fragile P.~C.,  Gillespie A.,  Monahan T.,  Rodriguez M.,    Anninos P.,  2012,
  \apjs, submitted

\bibitem[\protect\citeauthoryear{{Hawley}, {Guan} \& {Krolik}}{{Hawley}
  et~al.}{2011}]{hawley11}
{Hawley} J.~F.,  {Guan} X.,    {Krolik} J.~H.,  2011, ArXiv e-prints

\bibitem[\protect\citeauthoryear{{Hawley} \& {Krolik}}{{Hawley} \&
  {Krolik}}{2006}]{hawley06}
{Hawley} J.~F.,  {Krolik} J.~H.,  2006, \apj, 641, 103

\bibitem[\protect\citeauthoryear{{Igumenshchev}}{{Igumenshchev}}{2008}]{igumenshchev08}
{Igumenshchev} I.~V.,  2008, \apj, 677, 317

\bibitem[\protect\citeauthoryear{{Igumenshchev}}{{Igumenshchev}}{2009}]{igumenshchev09}
{Igumenshchev} I.~V.,  2009, \apjl, 702, L72

\bibitem[\protect\citeauthoryear{{Igumenshchev}, {Narayan} \&
  {Abramowicz}}{{Igumenshchev} et~al.}{2003}]{igumenshchev03}
{Igumenshchev} I.~V.,  {Narayan} R.,    {Abramowicz} M.~A.,  2003, \apj, 592,
  1042

\bibitem[\protect\citeauthoryear{{McClintock} \& {Remillard}}{{McClintock} \&
  {Remillard}}{2006}]{mcclintock06}
{McClintock} J.~E.,  {Remillard} R.~A.,  2006, {Black hole binaries}.
pp 157--213

\bibitem[\protect\citeauthoryear{{McKinney}}{{McKinney}}{2006}]{mckinney06}
{McKinney} J.~C.,  2006, \mnras, 368, 1561

\bibitem[\protect\citeauthoryear{{McKinney} \& {Blandford}}{{McKinney} \&
  {Blandford}}{2009}]{mckinney09}
{McKinney} J.~C.,  {Blandford} R.~D.,  2009, \mnras, 394, L126

\bibitem[\protect\citeauthoryear{{McKinney} \& {Gammie}}{{McKinney} \&
  {Gammie}}{2004}]{mckinney04}
{McKinney} J.~C.,  {Gammie} C.~F.,  2004, \apj, 611, 977

\bibitem[\protect\citeauthoryear{{McKinney}, {Tchekhovskoy} \&
  {Blandford}}{{McKinney} et~al.}{2012}]{mckinney12}
{McKinney} J.~C.,  {Tchekhovskoy} A.,    {Blandford} R.~D.,  2012, ArXiv
  e-prints

\bibitem[\protect\citeauthoryear{{Mo{\'s}cibrodzka} \&
  {Proga}}{{Mo{\'s}cibrodzka} \& {Proga}}{2009}]{moscibrodzka09}
{Mo{\'s}cibrodzka} M.,  {Proga} D.,  2009, \mnras, 397, 2087

\bibitem[\protect\citeauthoryear{{Narayan} \& {McClintock}}{{Narayan} \&
  {McClintock}}{2011}]{narayan11}
{Narayan} R.,  {McClintock} J.~E.,  2011, \mnras, p.~L362

\bibitem[\protect\citeauthoryear{{Narayan} \& {Yi}}{{Narayan} \&
  {Yi}}{1994}]{narayan94}
{Narayan} R.,  {Yi} I.,  1994, \apjl, 428, L13

\bibitem[\protect\citeauthoryear{{Noble}, {Krolik} \& {Hawley}}{{Noble}
  et~al.}{2009}]{noble09}
{Noble} S.~C.,  {Krolik} J.~H.,    {Hawley} J.~F.,  2009, \apj, 692, 411

\bibitem[\protect\citeauthoryear{{Ponti}, {Fender}, {Begelman}, {Dunn},
  {Neilsen} \& {Coriat}}{{Ponti} et~al.}{2012}]{ponti_12}
{Ponti} G.,  {Fender} R.~P.,  {Begelman} M.~C.,  {Dunn} R.~J.~H.,  {Neilsen}
  J.,    {Coriat} M.,  2012, ArXiv e-prints

\bibitem[\protect\citeauthoryear{{Russell}, {Miller-Jones}, {Maccarone},
  {Yang}, {Fender} \& {Lewis}}{{Russell} et~al.}{2011}]{russell11}
{Russell} D.~M.,  {Miller-Jones} J.~C.~A.,  {Maccarone} T.~J.,  {Yang} Y.~J.,
  {Fender} R.~P.,    {Lewis} F.,  2011, \apjl, 739, L19

\bibitem[\protect\citeauthoryear{{Shakura} \& {Sunyaev}}{{Shakura} \&
  {Sunyaev}}{1973}]{shakura73}
{Shakura} N.~I.,  {Sunyaev} R.~A.,  1973, \aap, 24, 337

\bibitem[\protect\citeauthoryear{{Tchekhovskoy} \& {McKinney}}{{Tchekhovskoy}
  \& {McKinney}}{2012}]{tchekhovskoy12}
{Tchekhovskoy} A.,  {McKinney} J.~C.,  2012, \mnras, p.~L445

\bibitem[\protect\citeauthoryear{{Tchekhovskoy}, {Narayan} \&
  {McKinney}}{{Tchekhovskoy} et~al.}{2011}]{tchekhovskoy11}
{Tchekhovskoy} A.,  {Narayan} R.,    {McKinney} J.~C.,  2011, \mnras, 418, L79

\bibitem[\protect\citeauthoryear{{Tomimatsu}}{{Tomimatsu}}{1994}]{tomimatsu94}
{Tomimatsu} A.,  1994, \pasj, 46, 123

\bibitem[\protect\citeauthoryear{{Wilms}, {Nowak}, {Pottschmidt}, {Pooley} \&
  {Fritz}}{{Wilms} et~al.}{2006}]{wilms06}
{Wilms} J.,  {Nowak} M.~A.,  {Pottschmidt} K.,  {Pooley} G.~G.,    {Fritz} S.,
  2006, \aap, 447, 245

\end{thebibliography}

\label{lastpage}

\end{document}